\documentclass[a4paper,12pt]{article}
\pdfoutput=1

\usepackage{jheppub}
\usepackage[utf8]{inputenc}
\usepackage[english]{babel}

\usepackage{amsfonts}
\usepackage{amsthm}
\usepackage{mathrsfs}
\usepackage[mathcal]{euscript}
\usepackage{float}
\newcommand{\be}{\begin{equation}}
\newcommand{\ee}{\end{equation}}

\newcommand{\skipline}{\vspace{\baselineskip}}

\newcommand{\nn}{\nonumber}
\newcommand{\eps}{\varepsilon}
\newcommand{\m}{_\text{m}}
\newcommand{\gen}{_\text{gen}}
\newcommand{\ext}{^\text{ext}}
\newcommand{\extgen}{\ext\gen}

\newcommand{\LateTimes}
{\Big\rvert_{\begin{subarray}{l}\text{late}\\\text{times}\end{subarray}}}
\newcommand{\BH}{_\text{B-H}}

\newcommand{\InterTimes}{\Big\rvert_{\begin{subarray}{l}\text{inter.}\\\text{times}\end{subarray}}}
\newcommand{\ba}{\mathbf{a}}
\newcommand{\bb}{\mathbf{b}}
\newcommand{\bq}{\mathbf{q}}

\newcommand{\bx}{\mathbf{x}}
\newcommand{\by}{\mathbf{y}}

\usepackage{comment}
\usepackage{soul}
\usepackage{xcolor}
\usepackage[normalem]{ulem}

\title{Finite entangling regions and information paradox in charged black holes}

\author{Aleksandr I. Belokon}

\affiliation{Steklov Mathematical Institute, Russian Academy of Sciences,\\ Gubkin str. 8, 119991 Moscow, Russian Federation}

\emailAdd{belokon@mi-ras.ru}

\abstract{In this paper, we study the influence of the electric charge of Reissner-Nordström black hole on the dynamics of fine-grained entropy of Hawking radiation, collected in finite entangling regions. We demonstrate that for certain sizes of finite regions, it is always possible to choose a value of the charge such that no information paradox formulated for finite entangling regions arises. For the sake of completeness, we explore how entanglement islands influence the described picture. We find that at small values of the electric charge, there is a discontinuity in the entropy due to the disappearance of the island, and with increasing charge the island ceases to ever dominate throughout entire evolution.}

\makeatletter
\gdef\@fpheader{\skipline}
\makeatother

\begin{document}
\maketitle
\flushbottom
\newpage

%-----------%

\section{Introduction}
Taking into account semiclassical effects at event horizons leads to a phenomenon called Hawking radiation~\cite{Hawking1975particle, Hawking:1976ra}. This phenomenon is the first and, in essence, so far the only manifestation of the effects of quantum gravity that can be formulated without having a complete theory describing it. The problem with Hawking radiation, called the information paradox, is that it is currently unclear how to reconcile its thermal nature with unitary evolution from a pure initial state. The evolution of the fine-grained entropy of Hawking radiation of an evaporating black hole is described by the Page curve~\cite{Page:1993wv, Page:2013dx}. In this context, the question of how to obtain its characteristic shape --- particularly its descendent part --- from explicit calculations still remains open.

Consideration of the problem of Hawking radiation in two-dimensional Jackiw-Teitelboim (JT) gravity led to the formulation of the island proposal widely discussed in the literature, which modifies the formula for the entropy in the presence of dynamical gravity. The island formula was originally known from holography~\cite{Ryu:2006bv, Hubeny:2007xt, Barrella:2013wja, Faulkner:2013ana, Engelhardt:2014gca}, and its alternative direct derivation in two dimensions was later obtained with the use of the replica trick in the gravitational path integral~\cite{Penington:2019npb, Almheiri:2019psf, Almheiri:2019hni}. 
Explicit calculations of entanglement islands are also done in some specific higher-dimensional models with holographic duals~\cite{Geng:2020qvw, Geng:2020fxl, Geng:2024xpj, Geng:2023qwm, Chen:2020uac, Rozali:2019day}, see also~\cite{Almheiri:2019psy}. In Karch-Randall braneworld models~\cite{Geng:2020qvw, Geng:2021hlu, Geng:2023qwm} and in JT gravity~\cite{Geng:2023zhq} the potential puzzle of inconsistency of the island proposal with the massless gravity is discussed. However, in the absence of a direct extension of the two-dimensional derivation in higher-dimensional geometries with horizons, we are currently forced to postulate that the main conclusions of the island proposal are also valid in them.

There are two general directions, in which attempts are being made to study the problem of Hawking radiation. In the first case, the original geometry is either initially two-dimensional dilaton gravity, or it is higher-dimensional and reduced to two dimensions via dimensional reduction. This geometry serves as the background for the action of two-dimensional conformal matter, modeling the Hawking quanta~\cite{Penington:2019kki, Almheiri:2019qdq, Almheiri:2019psf, Almheiri:2019hni,Hartman:2020swn,Anegawa:2020ezn,Djordjevic:2022qdk,Ahn:2021chg,Chen:2020tes, Hartman:2020khs, Balasubramanian:2020xqf,Aalsma:2021bit,Sybesma:2020fxg,Svesko:2022txo}. The second approach, which we will follow in this paper, uses the s-wave approximation proposed in~\cite{Penington:2019npb} and used in the context of higher-dimensional Schwarzschild black hole in~\cite{HIM}. This approximation, applied to the fields in the background of a higher-dimensional spherically-symmetric black hole, is supposed to effectively reduce the problem to a two-dimensional CFT. The variety of papers exploiting the s-wave approximation in different contexts have been published recently~\cite{Alishahiha:2020qza, Matsuo:2020ypv, Karananas:2020fwx, Yu:2021cgi, Ling:2020laa, Lu:2021gmv, Wang:2021woy, Ahn:2021chg, Arefeva:2021kfx, Stepanenko:2022gwy, Yu:2021rfg, Azarnia:2021uch, Cao:2021ujs, Kim:2021gzd, He:2021mst, Ageev:2022hqc, Arefeva:2022cam, Djordjevic:2022qdk, Azarnia:2022kmp, Gan:2022jay, Omidi:2021opl}.

Apart from the entangling regions of infinite extent, finite regions are of independent interest in the context of the information paradox. The first work to address the thermal bath of finite size studied this problem in the Karch-Randall braneworld model using the holographic duality~\cite{Geng:2021iyq}. In the paper~\cite{Ageev:2022qxv}, the fine-grained entropy of Hawking radiation collected in regions of finite size was studied in the context of the higher-dimensional Schwarzschild black hole. The latter model was motivated by the study of black holes in asymptotically de Sitter space, and the outer boundaries of finite entangling regions simulated the finiteness of the observable domain due to the presence of the cosmological horizon. However, the features of the entropy discovered in this setup turned out to be interesting by themselves and raised the question of formulating and resolving the information paradox in a new context.

In this paper, we will study finite regions, which in~\cite{Ageev:2022qxv} are classified as ``mirror-symmetric'' (see Fig.~\ref{fig:4}). Such entangling regions represent the union of two finite subregions in the static patches of an analytically extended black hole geometry. The boundaries of these subregions are located on the same time slices, realizing the situation when the radiation is collected in the domain between two concentric spheres of different radii. In the right static patch, the boundaries move along the corresponding timelike Killing vectors, while in the left patch --- in the opposite direction. Since the described picture is not an isometry, the overall problem becomes non-stationary~\cite{Almheiri:2019yqk, Hartman:2013qma}. 

The rate of growth of the entanglement entropy of conformal matter for such finite regions in Schwarzschild black hole at early times is twice the ``canonical'' rate calculated for semi-infinite regions (see the corresponding calculations in~\cite{HIM,Ageev:2022qxv}). At late times, the entropy reaches saturation, and its value increases for larger regions. For such regions, the island configurations exist only for a finite time~\cite{Ageev:2022qxv}. After the island disappears, there is a discontinuity in the entropy because during the island domination, the matter entropy grows twice as fast as the entropy with the island. The strong bound of entanglement entropy (see section~\ref{sec:IPFR}) becomes violated for a finite period, which is longer the larger the region. This is a clear manifestation of the information paradox for finite entangling regions. Therefore, the island proposal does not resolve the information paradox, at least in the described setup.

In this paper, we extend the approach described in~\cite{Ageev:2022qxv} to the case of Reissner-Nordström black hole. We consider the problem of the information paradox for finite regions and explore how the electric charge of the black hole influences the behavior of entanglement entropy, as well as how the island proposal affects the described picture. We show that for finite regions of sufficiently large extent it is always possible to choose such a minimum charge value, starting from which the information paradox does not arise. In turn, the island formula gives two main effects: at small values of the charge, it also leads to a discontinuity in the entropy, just like in the case of Schwarzschild black hole. As the charge increases, the island configuration is always subdominant and does not contribute to the final answer for the entropy. Thus, the island formula does not solve the information paradox for finite regions in Reissner-Nordström black hole as well. 

The paper is organized as follows. In section~\ref{sec:setup}, we give a brief overview of the setup. In section~\ref{sec:entmatter}, we study the features of entanglement entropy of matter for finite regions in the presence of an electric charge as well as the strong bound on entanglement entropy in Reissner-Nordström black hole. In section~\ref{sec:genentfunc}, we study how entanglement islands affect the evolution of finite regions and how it is changed when tending to the extremal case. In section~\ref{sec:discuss}, we briefly discuss the obtained results.

\section{Setup}\label{sec:setup}
\subsection{Geometry}
We start with the metric of the four-dimensional Reissner-Nordström black hole
\be
    ds^2 = -f(r)dt^2 + \frac{dr^2}{f(r)} + r^2 d\Omega_2^2, \qquad f(r) = 1 - \frac{2GM}{r} + \frac{GQ^2}{r^2},
    \label{eq:RN_metric}
\ee
where $M$ and $Q$ are the mass and the electric charge of the black hole, respectively, and $d\Omega_2^2$ is the angular part of the metric. The outer and the inner horizons are the roots of the blackening function
\be\label{eq:frprm}
    f(r) = \frac{1}{r^2}(r - r_+)(r - r_-), \qquad r_\pm \equiv GM \pm \sqrt{(GM)^2 - GQ^2}.
\ee
The Hawking temperature $T_H$ and the surface gravity $\kappa_h$ of this charged black hole are given by
\be\label{eq:T-H}
    T_H = \frac{r_+ - r_-}{4\pi r_+^2} = \frac{1}{2\pi}\frac{\sqrt{(GM)^2 - GQ^2}}{\left(GM + \sqrt{(GM)^2 - GQ^2}\right)^2},
\ee
\be\label{eq:kappa}
    \kappa_h = 2\pi T_H = \frac{r_+ - r_-}{2r_+^2}.
\ee

Let us introduce Kruskal coordinates,
which for the right static patch read
\be
    U = -\frac{1}{\kappa_h}\,e^{-\kappa_h(t - r_*)}, \qquad V = \frac{1}{\kappa_h}\,e^{\kappa_h(t + r_*)},
    \label{eq:right_wedge_Krusk}
\ee
where the tortoise coordinate $r_*(r)$ is defined as
\be
    r_*(r) = \int\limits^r\frac{dr}{f(r)} = r + \left(\frac{r_+^2}{r_- - r_+}\right)\ln\left|\frac{r - r_+}{r_+}\right| - \left(\frac{r_-^2}{r_+ - r_-}\right)\ln\left|\frac{r - r_-}{r_-}\right|.
\ee
In terms of these coordinates, we can write the metric~\eqref{eq:RN_metric} in the following form
\be
    ds^2 = -e^{2 \rho(r)} dU dV + r^2 d\Omega^2,
    \label{eq:krusk-metr}
\ee
with the conformal factor $e^{2 \rho(r)}$ given by
\be
    e^{2 \rho(r)} = f(r) e^{-2\kappa_h r_*}.
    \label{eq:conformalfactor}
\ee
The squared radial distance $d^2(\bx, \by)$ for the spherically symmetric two-dimensional part of the metric~\eqref{eq:krusk-metr} reads
\be
    d^2(\bx, \by) = e^{\rho(\bx)} e^{\rho(\by)}\left[U(\bx) - U(\by)\right] \left[V(\by) - V(\bx)\right],
    \label{eq:d2}
\ee
This formula can be derived if we consider the spherically-symmetric part of the metric in Kruskal coordinates~\eqref{eq:krusk-metr} as a Weyl transformed version of its flat counterpart ${ds^2 = -dU\,dV}$ with the Weyl factor $e^{2\rho(r)}$ given by~\eqref{eq:conformalfactor}. Bold letters denote pairs of static radial and time coordinates, e.g., $\bx = \left(x, t_x\right)$. In terms of $(t, r)$-coordinates, the distance~\eqref{eq:d2} is given by
\be
    d^2(\bx, \by) = 
        \frac{2 \sqrt{f(x)f(y)}}{\kappa^2_h} \big[\cosh \kappa_h (r_*(x) - r_*(y)) - \cosh\kappa_h (t_x - t_y)\big].
    \label{eq:geod_dist}
\ee

We use the following notation for spacetime points in the right and left wedges of the Penrose diagram for this black hole, respectively,
\be\nn
    \bx_+ = \left(x_+,\,t_{x_+}\right), \qquad \bx_- = \left(x_-,\,t_{x_-} + \frac{i \pi}{\kappa_h}\right).
    \label{eq:notation_points}
\ee
Note that the imaginary part of the time coordinate of $\bx_-$ implies that this point is in the left wedge.

\subsection{Entanglement entropy in higher-dimensional setups}
Calculation of entanglement entropy in a higher-dimensional curved spacetime is challenging. 
The main difficulties are as follows:
\begin{itemize}
    \item the gravitational formula for the entropy~\eqref{eq:Sgen} is rigorously derived only in holography~\cite{Ryu:2006bv, Hubeny:2007xt, Barrella:2013wja, Faulkner:2013ana, Engelhardt:2014gca} and in two-dimensional Jackiw-Teitelboim gravity~\cite{Penington:2019npb, Almheiri:2019psf, Almheiri:2019hni};

    \item the matter contribution $S\m$ to the entropy in the island formula~\eqref{eq:Sgen} can be calculated only in some special cases. One of the few such cases is the two-dimensional conformal field theory, in which the presence of conformal symmetry allows one to reduce the replica calculation of the Renyi entropy to the calculation of the two-point correlation function of special operators called twistor operators. The result of these calculations are the formulas given below: ~\eqref{eq:S_1_ints} and~\eqref{eq:S_N_ints}~\cite{Calabrese:2004eu, Calabrese:2009qy, Casini:2005rm}.
\end{itemize}

While the first point implies that we have only to postulate the island formula~\eqref{eq:Sgen} in higher dimensions, the second one is the reason why the problem should be reduced to a two-dimensional conformal field theory in one way or another. There are two main approaches in the literature to reducing a higher-dimensional problem to a two-dimensional conformal field theory: the s-wave approximation, which we use in this paper, i.e.
\be
    S_\text{pure geometry} + S_\text{matter} \xrightarrow[]{\text{s-wave}} S_\text{CFT$_2$},
\ee
and dimensional reduction of pure (matter-free) higher-dimensional geometry to two dimensions with consideration of two-dimensional conformal field theory on the resulting background, i.e.
\be
    S_\text{pure geometry} \xrightarrow[]{\text{dimensional reduction}} S_\text{reduced geometry} + S_\text{CFT$_2$}.
\ee
It is assumed that both approaches reflect the basic properties of the entanglement entropy of Hawking radiation in the original higher-dimensional problem. Generally speaking, they are not equivalent, but, for example, the study of the entanglement entropy in Schwarzschild black hole using the s-wave approximation~\cite{HIM} and dimensional reduction~\cite{Djordjevic:2022qdk} gave similar results.

The main idea, according to which the original higher-dimensional theory in the s-wave approximation is a two-dimensional conformal field theory, is as follows. Consideration of the Klein-Gordon equation in the background of a spherically symmetric asymptotically flat black hole leads to an effective ``massive'' term $\propto \ell(\ell + 1)\phi^2$, which vanishes in the case $\ell = 0$. This fact serves as a justification for that the s-mode is a conformal field. This statement, although it requires more rigorous proofs and can be disputed, is widely used in the literature~\cite{Penington:2019npb, HIM, Alishahiha:2020qza, Matsuo:2020ypv, Karananas:2020fwx, Yu:2021cgi, Ling:2020laa, Lu:2021gmv, Wang:2021woy, Ahn:2021chg, Arefeva:2021kfx, Yu:2021rfg, Azarnia:2021uch, Cao:2021ujs, Kim:2021gzd, He:2021mst, Ageev:2022hqc, Arefeva:2022cam, Djordjevic:2022qdk, Anand:2022mla, Azarnia:2022kmp, Gan:2022jay}. In addition, the effective potential of the scalar field in the geometry of such a black hole acts as a barrier, which is the smallest for the mode with $\ell = 0$, i.e. the observer at spatial infinity collects predominantly s-modes, which justifies the replacement of the entire spectrum of Hawking radiation by only its s-wave part.

After reducing the original problem to a two-dimensional conformal field theory, we can use the well-known expressions for the entanglement entropy for one interval and $N \geq 1$ disjoint intervals, respectively,
\be
    S\m = \frac{c}{3} \ln\frac{d(\bx, \by)}{\eps},
    \label{eq:S_1_ints}
\ee
\be
    S\m = \frac{c}{3} \sum_{i,\,j} \ln\frac{d(\bx_{i}, \by_{j})}{\eps} - \frac{c}{3} \sum_{i\,<\,j}^{N}\ln\frac{d(\bx_{i}, \bx_{j})}{\eps} - \frac{c}{3} \sum_{i\,<\,j} \ln\frac{d(\by_{i}, \by_{j})}{\eps},
    \label{eq:S_N_ints}
\ee
where the distance $d(\bx_i,\by_j)$ is given by~\eqref{eq:geod_dist}, $\bx_i$ and $\by_i$ denote left and right endpoints of the corresponding intervals, and $\eps$ is a UV cutoff. Note that while the formula~\eqref{eq:S_1_ints} is valid in an arbitrary CFT$_2$, the expression~\eqref{eq:S_N_ints} describes the entanglement entropy of $c$ free massless Dirac fermions\footnote{This formula for Dirac fermions in flat space was derived in~\cite{Casini:2005rm}. Considering that any two-dimensional metric is conformally flat and using the transformation properties of entanglement entropy under Weyl transformations~\cite{Almheiri:2019psf}, one can derive the formula~\eqref{eq:S_N_ints} for the curved background~\eqref{eq:krusk-metr} at fixed values of angular coordinates. The entanglement entropy of free massless Dirac fermions in a curved background has been considered in the context of the island proposal~\cite{Almheiri:2019qdq,Azarnia:2021uch, Kawabata:2021vyo} and for inhomogeneous non-interacting Fermi gases \cite{Dubail:2016tsc}.}. Since in the original higher-dimensional problem the union $R \cup I$ of the region $R$ where the radiation is collected and the entanglement island $I$ is a disconnected domain, we need a formula for the entanglement entropy of the union of $N$ disconnected intervals in a two-dimensional conformal field theory. Since this formula is known only for a two-dimensional conformal field theory represented by free massless Dirac fermions~\cite{Casini:2005rm}, we will further assume that the Hawking quanta are represented by precisely this type of matter.

\subsection{Entanglement entropy in the presence of gravity}
Recently, it was shown that in some gravitational setups the unitary behavior of the Page curve can be derived using the island proposal~\cite{Cotler:2017erl, Almheiri:2019hni, Penington:2019npb, Almheiri:2019qdq,Penington:2019kki}. The proposal goes as follows: first, we are to consider the so-called generalized entropy funtional, which reads
\be
    S\gen[I, R] = \frac{\operatorname{Area}\,(\partial I)}{4G} + S\m(R \cup I).
    \label{eq:gen_functional}
\ee
Here $\partial I$ denotes the boundary of the entanglement island, and $S\m$ is the entanglement entropy of conformal matter. Then we should extremize this functional over all possible island configurations
\be
    S\extgen[I, R] = \underset{\partial I}{\operatorname{ext}}\,\Big\{S\gen[I, R]\Big\},
\ee
and then choose the minimal one
\be
    S(R) = \underset{\partial I}{\text{min}}\,\Big\{S\extgen[I, R]\Big\}.
    \label{eq:Sgen}
\ee
The latter expression is called the island formula. In the following, we will use this formula along with the expressions~\eqref{eq:S_1_ints} and~\eqref{eq:S_N_ints} for the entanglement entropy of conformal matter $S\m$.

\subsection{Information paradox for finite regions}\label{sec:IPFR}
Let us sketch briefly the formulation of the information paradox for finite entangling regions in a black hole geometry. Similar discussion can be found in~\cite{Ageev:2022qxv}. 

Let us divide a Cauchy surface in a two-sided geometry into the region associated with the ``black hole system''~$BH$, a finite entangling region~$R$ and an adjacent semi-infinite region~$C$, which extends to spacelike infinities $i^0$
\be
    \Sigma = BH \cup R \cup C.
    \label{eq:tripartition}
\ee
We should emphasize here that by the ``black hole system'' we mean the domain extending between the nearest (to the black hole horizons) boundaries of the region $R$ (see Fig.~\ref{fig:4}, left), which play the role of the cutoff surfaces~\cite{Almheiri:2020cfm, Raju:2020smc}. Roughly, the cutoff surface divides the initial manifold into the ``black hole geometry'' and the ``outside geometry'', where in the latter gravity is set to be ``weak'' in the sense that this geometry is considered as Minkowski space. 

The strong subadditivity of entanglement entropy~\cite{Nishioka:2018khk} for a tripartition like~\eqref{eq:tripartition} gives the inequality
\be
    S(BH \cup R \cup C) + S(R) \leq S(BH \cup R) + S(R \cup C).
\ee
Then, using the complementarity property
\be
    \begin{aligned}
        & S(BH) = S(R \cup C), \\
        & S(R) = S(BH \cup C), \\
        & S(C) = S(BH \cup R),
    \end{aligned}
    \label{eq:equalityoftripart}
\ee
and pure state condition for the total state, $S(\Sigma) = 0$, we derive the upper bound on the entanglement entropy for a finite region~$R$, which is called \textit{the strong bound}
\be
    S(R) \leq 2 S\BH + S(C),
    \label{eq:IPFinite}
\ee
where 
\be\label{eq:SBH}
    S\BH = \frac{\text{Area$\,$(horizon)}}{4G},
\ee
is Bekenstein-Hawking entropy. The violation of this bound can be seen as the information paradox for finite entangling regions. 
In turn, the island proposal~\eqref{eq:Sgen} leads to softening of this constraint --- \textit{the soft bound}
\be
    S(R) \leq 2 S\BH + S(C) + S_\text{corr}.
    \label{eq:IPFinite_soft}
\ee
The correction $S_\text{corr}$ is time-independent and small compared to the area term $S\BH$ under the ``black hole classicality'' condition~\cite{HIM}
\be
    \frac{r^2_h}{G} \gg c.
    \label{eq:BH-classicality}
\ee
In other words, we say that the information paradox in two-sided black hole does not arise if either the bound~\eqref{eq:IPFinite} is respected or violated only by terms suppressed under the condition~\eqref{eq:BH-classicality}.

\section{Entropy of matter}\label{sec:entmatter}
\subsection{Finite region $R$}
\begin{figure}[ht!]\centering
    \includegraphics[width=0.46\textwidth]{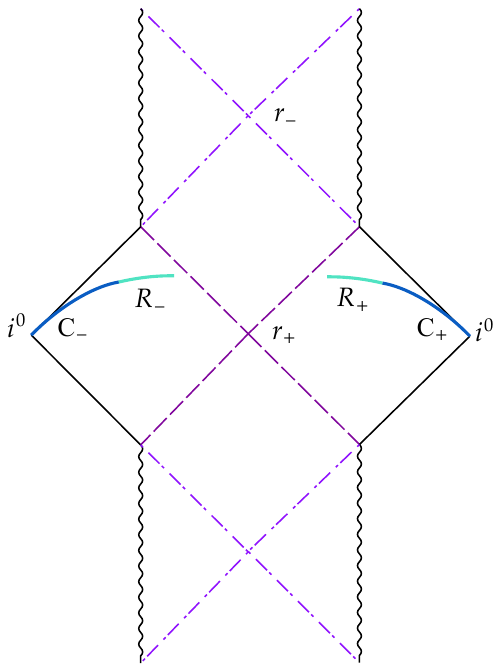} \hspace{0.05\textwidth}
    \includegraphics[width=0.46\textwidth]{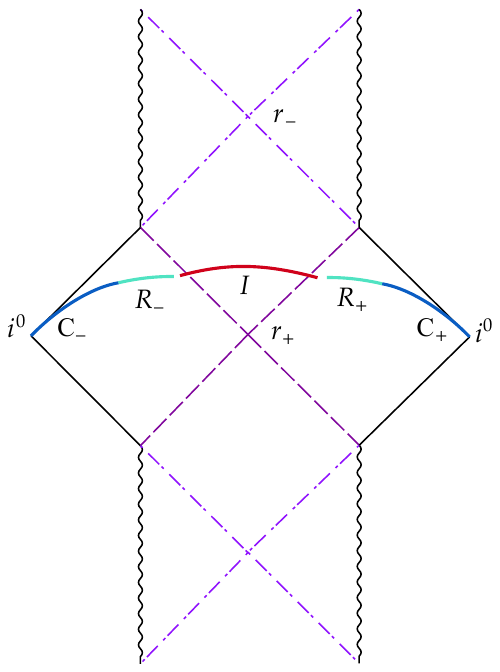}
	\caption{Part of the Penrose diagram of Reissner-Nordström spacetime, with the finite region $R = R_- \cup R_+$~\eqref{eq:RMS} (light blue) and the semi-infinite adjacent region $C = C_- \cup C_+$~\eqref{eq:CMS} (blue) indicated. The outer horizon $r_+$ is shown by dashed lines, and the inner horizon $r_-$ --- by dotted-dashed ones. \emph{Left:} The case without island. \emph{Right:} The case with the island~$I$~\eqref{eq:isl} for the region $R$ (red).}
    \label{fig:4}
\end{figure}

Consider a finite region $R$ with the boundaries given by (see Fig.~\ref{fig:4})
\be\label{eq:RMS}
    R = R_- \cup R_+ \equiv [\bq_{-},\, \bb_{-}] \cup [\bb_{+},\, \bq_{+}],
\ee
where
\be\label{eq:RMSbq}
    \bq_- = \left\{q, -t_b + \frac{i\pi}{\kappa_h}\right\},  \quad \bq_+ = \left\{q, t_b\right\},
\ee
\be\nn
    \bb_- = \left\{b, -t_b + \frac{i\pi}{\kappa_h}\right\},  \quad \bb_+ = \left\{q, t_b\right\}.
\ee
With respect to a static observer, one can imagine each region $R_\pm$ as a domain between two concentric spheres with radii $b$ and $q > b$. Outgoing Hawking modes pass through this domain in a finite time and then escape to infinity.

\begin{figure}[t!]\centering
    \includegraphics[width=0.46\textwidth]{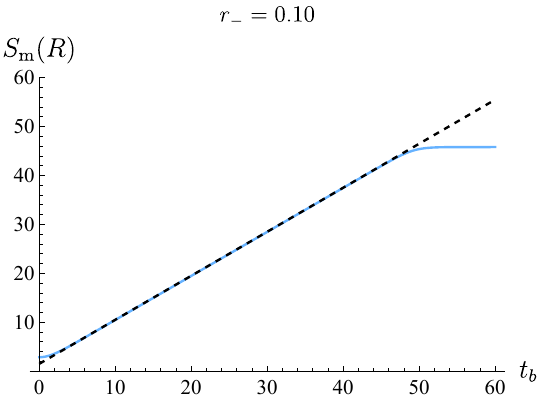} \hspace{0.05\textwidth}
    \includegraphics[width=0.46\textwidth]{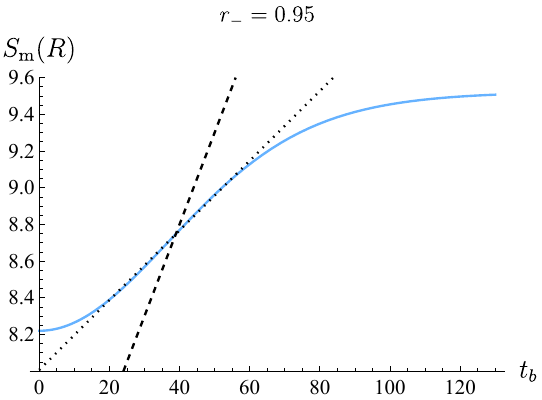}
	\caption{Entanglement entropy evolution of the region $R$~\eqref{eq:RMS} with $b = 5$ and $q = 100$ for different $r_-$. The linear-growth regime~\eqref{eq:mongrowthUP} is denoted by dashed lines, while the dotted line (right plot) denotes the actual rate of growth of the entropy near extremality. The entropy growth gets slower as $r_-$ increases. We take the parameters as $r_+ = 1$, $c = 3$, $G = 0.1$, $\eps = 1$.}
    \label{fig:1}
\end{figure}

The entanglement entropy for this region is given by
\be\label{eq:SRMS}
    S\m(R) = \frac{c}{3}\ln\left(\frac{4\sqrt{f(b)f(q)}}{\kappa^2_h \eps^2}\right) + \frac{2c}{3}\ln\cosh\kappa_h t_b + \frac{c}{3}\ln\left(\frac{\cosh\kappa_h(r_*(q) - r_*(b)) - 1}{\cosh\kappa_h(r_*(q) - r_*(b)) + \cosh 2\kappa_h t_b}\right).
\ee
The first and the third terms describe finite-size effects. The second term is responsible for the linear-growth regime at intermediate times. The dynamics of the entropy far from extremality ($r_- \not\to r_+$ and $\kappa_h \not\to 0$) is as follows (see Fig.~\ref{fig:1}):
\begin{itemize}
    \item at intermediate times, $1 \ll \cosh 2\kappa_h t_b \ll \cosh \kappa_h (r_*(q) - r_*(b))$ (see the third term in~\eqref{eq:SRMS}),
    the entanglement entropy of the radiation increases monotonically
    \be
        S\m(R)\InterTimes \simeq \frac{2 c}{3}\,\kappa_h t_b.
        \label{eq:mongrowthUP}
    \ee
    This increase is twice as fast as for the semi-infinite region with the same configuration of the boundaries $\bb$ (i.e., when the coordinates of $\bb$ are given by~\eqref{eq:RMSbq}, and the exterior boundaries $\bq$ are sent to the corresponding spatial infinities $i^0$.).

    \item At late times, $\cosh 2\kappa_h t_b \gg \cosh\kappa_h(r_*(q) - r_*(b))$ (when the finiteness of the region becomes important), the time-dependent terms in~\eqref{eq:SRMS} mutually cancel, and
    the entropy saturates at a constant value, which reads
    \be
        S\m(R)\LateTimes \simeq \frac{c}{3}\ln\left(\frac{2\sqrt{f(b)f(q)}}{\kappa_h^2\eps^2}\right) + \frac{c}{3}\ln\Big(\cosh\kappa_h(r_*(q) - r_*(b)) - 1\Big).
        \label{eq:entropyfinitenoisland}
    \ee
    The moment, at which the entropy reaches a constant value, is approximately equal to
    \be\label{eq:t-sat}
        t_\text{sat} \simeq \frac{r_*(q) - r_*(b)}{2}.
    \ee
Therefore, the larger the size of the region, the later the entropy reaches saturation. Also, a decrease in the size of the region $R$ leads to a faster onset of the moment of saturation, which results in a decrease in the final entropy value.
\end{itemize}

This time evolution of the entropy for the region $R$~\eqref{eq:RMS} away from extremality coincides with the evolution of the same region in Schwarzschild black hole~\cite{Ageev:2022qxv}. However, the fact that in our case the black hole carries an electric charge affects the entropy dynamics in two ways. Namely, the closer the black hole is to extremality ($r_- \to r_+$, $\kappa_h \to 0$):
\begin{itemize}
    \item the slower the rate of the entropy growth;

    \item the longer the entropy reaches saturation.
\end{itemize}

The entropy growth~\eqref{eq:mongrowthUP} gets slower, since it is determined by the surface gravity~$\kappa_h$~\eqref{eq:kappa} of the black hole, which, together with the Hawking temperature $T_H$~\eqref{eq:T-H}, decreases as the black hole approaches extremality. Moreover, near extremality the linear-growth regime is no longer described by the formula~\eqref{eq:mongrowthUP}, see Fig.~\ref{fig:1}. To obtain the correct formula for the rate of growth of the entropy near extremality, which can be parameterized as the limit $r_- = r_+ - \epsilon$, $\epsilon \to 0$, let us expand the entropy~\eqref{eq:SRMS} near the inflection point $t_\text{inf}$ and at small $\epsilon$
\be
    S\m(R) \simeq \frac{\epsilon^4 c \, t_\text{inf}}{96r_-^8} \big(r_*(b) - r_*(q)\big)^2 \cdot t_b + O\left(\epsilon^5\right),
\ee
where $t_\text{inf}$ is the solution to the equation $d^2 S\m(R)/dt_b^2 = 0$. We see that near extremality the linear growth is highly suppressed.

To illustrate the second statement about the time of saturation, let us consider the time-dependent part of the entropy~\eqref{eq:SRMS} near extremality
\be
    \frac{c}{3}\ln\left(\frac{\cosh^2\kappa_h t_b}{\cosh\kappa_h(r_*(q) - r_*(b)) + \cosh 2\kappa_h t_b}\right).
\ee
The saturation happens when this logarithm becomes zero and the entropy reaches a constant value, which does not depend on time $t_b$. Note that the expression under the logarithm in the $\epsilon \to 0$ limits does not depend on $t_b$ up to the 4$^\text{th}$ order in $\epsilon$. Expanding this expression about $\epsilon = 0$ and solving the equation $\ln(...) = 0$, we obtain in the leading order the solution for $t_{\text{ext,sat}}$
\be\label{eq:textsat}
    t_{\text{ext,sat}} \simeq \frac{8r_-^4}{\epsilon^2 \big(r_*(q) - r_*(b)\big)}.
\ee
Therefore, the time of saturation increases significantly when approaching the extremal case and formally diverges when $\epsilon = 0$.

\subsection{Adjacent semi-infinite region $C$ and the strong bound on entanglement entropy}
Now let us consider the adjacent (to the finite region $R$~\eqref{eq:RMS}, see Fig.~\ref{fig:4}) region $C$
\be\label{eq:CMS}
    C = C_- \cup C_+ \equiv [i^0,\, \bq_{-}] \cup [\bq_{+},\, i^0].
\ee
The entropy for this region is given by
\be
    S\m(C) = \frac{c}{6}\ln\left(\frac{4f(q)}{\kappa_h^2\eps^2}\cosh^2\kappa_h t_q\right).
\ee

\iffalse
Time evolution of $S(C)$ compared to the one of $S(R)$ is depicted on Fig.~\ref{fig:2}. 

\begin{figure}[t!]\centering
    \includegraphics[width=0.46\textwidth]{images/q100SCrmDependenceNoIsl1.pdf} \hspace{0.05\textwidth}
    \includegraphics[width=0.46\textwidth]{images/q100SCrmDependenceNoIsl2.pdf}
	\caption{\emph{Left:} Entanglement entropy evolution of the adjacent semi-infinite region $C$ (dashed) with $q = 100$ compared to the evolution of the entropy for the finite region $R$ (solid) with $b = 5$ and $q = 100$ for different $r_-$. \emph{Right:} The same evolution near extremality $r_- \to r_+$. We take the parameters as $r_+ = 1$, $c = 3$, $G = 0.1$, $\eps = 1$.}
    \label{fig:2}
\end{figure}
\fi

\begin{figure}[ht!]\centering
    \includegraphics[width=0.6\textwidth]{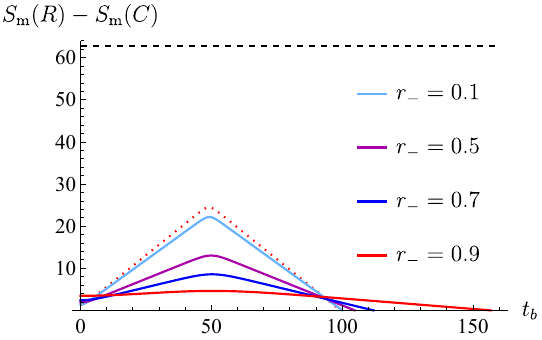} \hspace{0.05\textwidth}
    \includegraphics[width=0.6\textwidth]{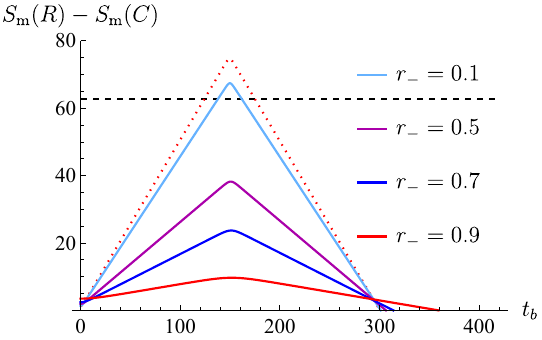}
    \caption{Time evolution of the difference between the entropies for finite region $R$~\eqref{eq:RMS} and its adjacent semi-infinite region $C$~\eqref{eq:CMS}, $S\m(R) - S\m(C)$, in the context of the strong bound on entanglement entropy of Hawking radiation~\eqref{eq:IPFinite} in Reissner-Nordström black hole with different $r_-$. The red dotted curves denote the same for Schwarzschild black hole ($r_- = 0$), and the black dashed lines depict twice the Bekenstein-Hawking entropy, $2S_\text{B-H}$. On the top figure, the finite region has $b = 5$ and $q = 100$, while on the bottom --- $b = 5$ and $q = 300$. We take the parameters as $r_+ = 1$, $c = 3$, $G = 0.1$, $\eps = 1$.}
    \label{fig:3}
\end{figure}

At late times, $\kappa_h t_b \gg 1$, the entropy $S(C)$ grows linearly. Since the rate of growth is determined by $\kappa_h$, near extremality the entropy increases slowly. Taking into account that the entropy $S(R)$ saturates at late times, see~\eqref{eq:entropyfinitenoisland}, and its final value is smaller the closer the black hole is to extremality, we can deduce that the strong bound on the entropy~\eqref{eq:IPFinite} might be satisfied starting from some $r_-$ for a given size of the region $q - b$, see Fig.~\ref{fig:3}. The value of the critical radius of the inner horizon $r_{-,\text{bound}}(q)$, starting from which the strong bound~\eqref{eq:IPFinite} is obeyed, can be found from
\be
    \frac{d}{dt_b}\left(S\m(R) - S\m(C)\right)\Big|_{t_b = t_{b,\text{max}}} = 0 \quad \implies \quad 
\ee
\be\label{eq:numsol}
    S\m(R) - S\m(C)\Big|_{t_b = t_{b,\text{max}}} = \frac{c}{6}\ln\left(\frac{f(b)}{\kappa_h^2\eps^2}\sinh^2\left(\frac{\kappa_h(r_*(q) - r_*(b))}{2}\right)\right) \leq 2S\BH.
\ee

\begin{figure}[ht!]\centering
    \includegraphics[width=0.55\textwidth]{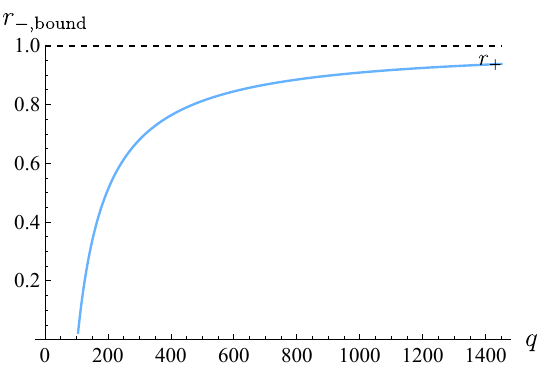} 
    \caption{Numerical solution to the inequality~\eqref{eq:numsol} with respect to the inner horizon $r_-(q)$, at which the strong bound~\eqref{eq:IPFinite} is satisfied. Lower values of $q$ correspond to negative solutions $r_-(q) < 0$, so that they are not considered. We take the parameters as $b = 5$, $r_+ = 1$, $c = 3$.}
    \label{fig:rmq}
\end{figure}

The numerical solution to the inequality~\eqref{eq:numsol} for $r_- = r_-(q)$ is presented in Fig.~\ref{fig:rmq}. We see that the larger the region, the closer to the extremal case $r_- \to r_+$ the black hole should be in order to avoid the information paradox for finite regions. Note that a positive solution $r_-(q) > 0$ exists not for any given $q > b > r_+$, but only starting from a certain value, which satisfies the condition $q \gg b > r_+$.

\section{Generalized entropy functional}\label{sec:genentfunc}
In Schwarzschild black hole the strong bound~\eqref{eq:IPFinite} can be satisfied only for small enough regions. Even though considering the generalized entropy functional can heal the problem up to corrections suppressed in the semiclassical limit~\eqref{eq:BH-classicality}, however, it can be shown that for finite regions of the type~\eqref{eq:RMS} entanglement islands always disappear at the moment when the entropy reaches saturation~\cite{Ageev:2022qxv,Ageev:2023hxe}. At this point, the entropy undergoes a discontinuous transition and disobeys the strong bound~\eqref{eq:IPFinite} for a finite period, which is longer the larger the region. In Reissner-Nordström black hole with small electric charges, $r_- \ll r_+$, the entropy experiences the same pattern of evolution with discontinuous transition. In this regard, the natural question arises on the influence of entanglement islands on the time evolution of the entropy in Reissner-Nordström geometry with $r_- \sim r_+$ given that for some finite regions and for some $r_-(q)$ the strong bound is satisfied automatically.

We consider the following symmetric ansatz for the entanglement island ${I = [\ba_-,\,\ba_+]}$ with
\be\label{eq:isl}
    \ba_- = \left(a, -t_a + \frac{i\pi}{\kappa_h}\right), \qquad \ba_+ = (a, t_a),
\ee
which is dictated by the symmetry of the region $R$~\eqref{eq:RMS}. With this ansatz the generalized entropy functional~\eqref{eq:gen_functional} takes the form
\be
    \begin{aligned}
    S\gen[I, R] & = \frac{2\pi a^2}{G} + \frac{c}{6}\ln\left(\frac{64f(a)f(b)f(q)}{\kappa_h^6 \eps^6}\right) + \frac{c}{6}\ln\cosh^2 \kappa_h t_a \cosh^4 \kappa_h t_b + \\
    & + \frac{c}{3}\ln\left(\frac{\cosh\kappa_h (r_*(q) - r_*(b)) - 1}{\cosh\kappa_h (r_*(q) - r_*(b)) + \cosh2\kappa_h t_b}\right) + \\
    & + \frac{c}{3}\ln\left(\frac{\cosh\kappa_h(r_*(b) - r_*(a)) - \cosh\kappa_h(t_b - t_a)}{\cosh\kappa_h (r_*(b) - r_*(a)) + \cosh\kappa_h(t_a + t_b)}\right) + \\
    & + \frac{c}{6}\ln\left(\frac{\cosh\kappa_h(r_*(q) - r_*(a)) + \cosh\kappa_h(t_a + t_b)}{\cosh\kappa_h (r_*(q)-r_*(a)) - \cosh\kappa_h(t_a-t_b)}\right).
    \end{aligned}
\ee

At intermediate times
\be
    \cosh\kappa_h (r_*(b) - r_*(a)) \ll \cosh\kappa_h(t_a + t_b) \ll \cosh\kappa_h (r_*(q) - r_*(a)),
    \label{eq:conditiononislandUP}
\ee
when the finiteness of the region does not play a role, there is an island solution
\be
    0 < a - r_+ \ll r_+.
\ee
At the moment~\eqref{eq:t-sat} the island disappears.

\begin{figure}[h!]\centering
    \includegraphics[width=0.6\textwidth]{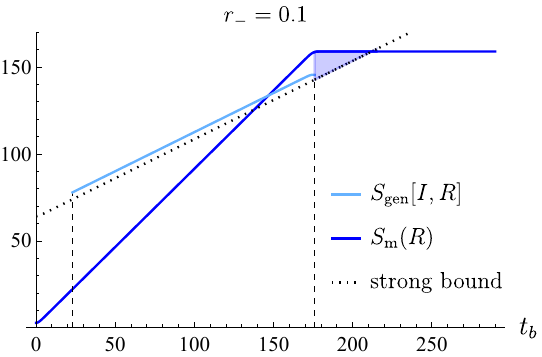} \hspace{0.05\textwidth}
    \includegraphics[width=0.6\textwidth]{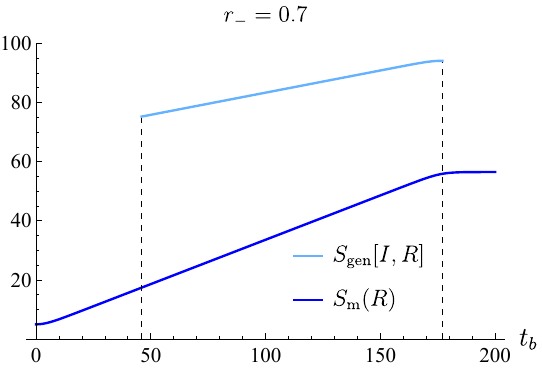}
    \caption{\emph{Top}: Time evolution of the entropy of matter (blue) and the generalized entropy functional with the island~\eqref{eq:isl} (sky blue) for a finite region $R$~\eqref{eq:RMS} in Reissner-Nordström black hole with $r_- = 0.1$. \emph{Bottom:} The same with $r_- = 0.7$. At each moment of time the value of the entropy is given by the minimum of these two curves.  When ${r_- \ll r_+}$, there is a discontinuity at the moment when the entropy of matter reaches a constant value (top plot), because the island disappears. The strong bound~\eqref{eq:IPFinite} (dotted line in the top plot) becomes violated over some finite period of time (shaded blue region in the top plot). Starting from some $r_-$, the dominant contribution comes from the entropy of matter $S\m(R)$ (lower plot). We take the parameters as $r_+ = 1$, $b = 5$, $q = 350$, $c = 3$, $G = 0.1$.}
    \label{fig:rm0107}
\end{figure}

Numerical analysis demonstrates that the island contributes only at small charges, $r_- \ll r_+$, and leads to a discontinuity in the entropy. When the island dominates, the entropy exceeds the strong bound within the correction, which satisfies the condition~\eqref{eq:BH-classicality}, thus, the information paradox does not arise in the sense of the soft bound~\eqref{eq:IPFinite_soft}, see Fig.~\ref{fig:rm0107}. As the black hole gets closer to the extremality, $r_- \to r_+$, the contribution of the island becomes subdominant and does not influence the final value of the entropy. Thus, the strong bound~\eqref{eq:IPFinite} in Reissner-Nordström black hole is obeyed starting from $r_{-,\text{bound}}$ for a given $q$, regardless whether we consider islands or not.

%%%%%%%%%%%%%%%%%%%%%%%%%%%%%%%%%

\section{Discussion}\label{sec:discuss}

\begin{figure}[th!]\centering
    \includegraphics[width=0.49\textwidth]{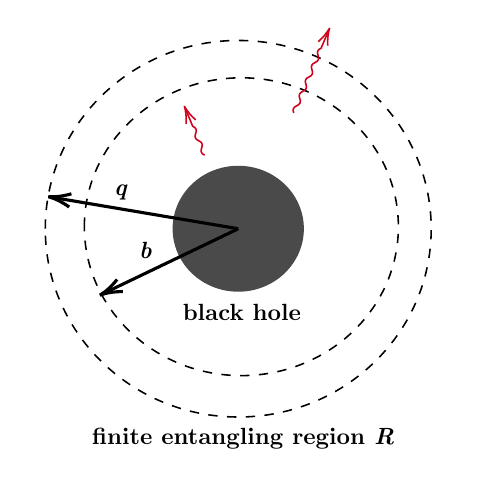} 
    \caption{Schematic picture of Hawking radiation, collected in a finite entangling region $R$~\eqref{eq:RMS}, whose boundaries are denoted by dashed concentric circles.}
    \label{fig:schematic}
\end{figure}

In this paper, we have demonstrated that the electric charge of a black hole significantly influences the entanglement entropy dynamics of Hawking radiation. The fact that the information paradox~\eqref{eq:IPFinite} does not arise for a sufficiently large class of finite regions when the black hole has an electric charge can be explained using the following reasoning. Let us consider the region $R$~\eqref{eq:RMS}, which, as we have indicated above, represents a spherical layer between two concentric spheres of radii $b$ and $q$ (see Fig.~\ref{fig:schematic}). As it was mentioned in~\cite{Ageev:2022qxv}, the saturation of the fine-grained entropy $S\m(R)$ at some finite value can be explained by the fact that from the moment the “first” Hawking quanta pass through the inner spherical boundary and until they cross the outer one, the entropy of such a region increases due to an increase in the number of quanta between them. As soon as the inward and outward flows become equal, the entropy takes on a constant value, which depends on the size of the region~$R$. Since Hawking radiation is thermal, its thermodynamic entropy can be derived from the Stefan-Boltzmann law and reads
\be
    S_\text{thermo}(R) \propto V(R) T_H^3,
\ee
where $V(R)$ is the volume of the region $R$, and $T_H$ is the Hawking temperature~\eqref{eq:T-H}. This entropy limits from above the fine-grained entropy $S\m(R)$ and, in turn, should not exceed the Bekenstein-Hawking entropy of the black hole $S\BH$~\eqref{eq:SBH}. Consequently, we obtain the following chain of inequalities
\be
    S\m(R) \leq S_\text{thermo}(R) \leq S\BH. 
\ee

In the case of Schwarzschild black hole, this inequalities explain why for sufficiently small regions ($q \gtrsim b$) the information paradox does not arise --- the thermodynamic entropy is proportional to the volume of the region, which can be made small enough so the strong bound~\eqref{eq:IPFinite} is satisfied. In the case of Reissner-Nordström black hole, as it tends to extremality, its temperature $T_H$ goes to zero, which reduces the thermodynamic entropy $S_\text{thermo}(R)$, while the value of Bekenstein-Hawking entropy $S\BH$ does not change significantly ($r_+ \to r_h/2$, where $r_h$ is the Schwarzschild radius). Thus, by choosing a sufficiently low temperature (or, equivalently, sufficiently large electric charge), we can ensure that the strong bound~\eqref{eq:IPFinite} is obeyed without involving the island formula~\eqref{eq:Sgen}. We believe that it is precisely this physical picture that our explicit calculations illustrate. 

\section*{Acknowledgements}
The author thanks Irina Ya. Aref'eva and Dmitry S. Ageev for posing the problem and valuable comments on the draft. The author would also like to express gratitude to Timofei A. Rusalev for fruitful discussions and valuable comments during the calculations. The study of the influence of the black hole electric charge on the dynamics of the entropy of Hawking radiation (sections 3 and 5) was supported by the Foundation for the Advancement of Theoretical and Mathematical Physics ``BASIS''. The study of the influence of entanglement islands on the dynamics of the entropy (sections 2 and 4) was supported by grant 24-72-10061 of the Russian Science Foundation.

%%%%%%%%%%%%%%%%%%%%

\bibliography{IslandsRN}
\bibliographystyle{JHEP}

\end{document}